\documentclass[prd,superscriptaddress,amsfonts,amssymb,amsmath,showpacs,twocolumn]{revtex4-2}
\usepackage{bm}
\usepackage{amsfonts}
\usepackage{latexsym}
\usepackage[latin1]{inputenc}
\usepackage{graphicx}
\usepackage{amsmath}
\usepackage{palatino}
\usepackage{mathpazo}
\usepackage{textcomp}
\usepackage{float}
\usepackage{booktabs}
\usepackage{dcolumn}
\usepackage{multirow}
\usepackage{ragged2e}
\usepackage{hyperref}
\hypersetup{colorlinks,citecolor=blue}
\usepackage{amsmath}
\usepackage{xcolor}
\usepackage{orcidlink}
\usepackage[caption=false]{subfig}
\usepackage{commath}
\def\jnl@style{\it}
\def\aaref@jnl#1{{\jnl@style#1}}

\def\aaref@jnl#1{{\jnl@style#1}}

\def\aj{\aaref@jnl{AJ}}                   
\def\apj{\aaref@jnl{ApJ}}                 
\def\apjl{\aaref@jnl{ApJ}}                
\def\apjs{\aaref@jnl{ApJS}}               
\def\apss{\aaref@jnl{Ap\&SS}}             
\def\aap{\aaref@jnl{A\&A}}                
\def\aapr{\aaref@jnl{A\&A~Rev.}}          
\def\aaps{\aaref@jnl{A\&AS}}              
\def\mnras{\aaref@jnl{Mon.~Not.~Roy.~Astron.~Soc.}}             
\def\prd{\aaref@jnl{Phys.~Rev.~D}}        
\def\prc{\aaref@jnl{Phys.~Rev.~C}}  
\def\prl{\aaref@jnl{Phys.~Rev.~Lett.}}    
\def\qjras{\aaref@jnl{QJRAS}}             
\def\skytel{\aaref@jnl{S\&T}}             
\def\ssr{\aaref@jnl{Space~Sci.~Rev.}}     
\def\zap{\aaref@jnl{ZAp}}                 
\def\nat{\aaref@jnl{Nature}}              
\def\aplett{\aaref@jnl{Astrophys.~Lett.}} 
\def\apspr{\aaref@jnl{Astrophys.~Space~Phys.~Res.}} 
\def\physrep{\aaref@jnl{Phys.~Rep.}}      
\def\physscr{\aaref@jnl{Phys.~Scr}}       
\def\commat{\aaref@jnl{Comm.~Math.~Phys.}}              
\def\science{\aaref@jnl{Science}}               
\def\cqg{\aaref@jnl{Classical Quant.~Grav.}}            
\def\jpcs{\aaref@jnl{JPCS}}                                     
\def\ijmpd{\aaref@jnl{Int.~J.~Mod.~Phys.~D}}                    
\def\grg{\aaref@jnl{Gen.~Relat.~Gravit.}}               
\def\rpp{\aaref@jnl{Rep.~Prog.~Phys.}}          
\def\npa{\aaref@jnl{Nucl.~Phys.~A}}        
\def\lrr{\aaref@jnl{Living Rev.~Rel.}}                   
\def\jcap{\aaref@jnl{J.~Cosmology Astropart.~Phys.}}    
\def\rmp{\aaref@jnl{Rev.~Mod.~Phys.}}   
\def\epjc{\aaref@jnl{Eur.~Phys.~J.~C}}

\allowdisplaybreaks[1]

\begin{document}

\color{black}       
\title{Cosmology in energy-momentum squared symmetric teleparallel gravity}

\author{Aaqid Bhat\orcidlink{0000-0002-8644-0677}}
\email{aaqid555@gmail.com}
\affiliation{Department of Mathematics, Birla Institute of Technology and
Science-Pilani,\\ Hyderabad Campus, Hyderabad-500078, India.}
\author{P.K. Sahoo\orcidlink{0000-0003-2130-8832}}
\email{pksahoo@hyderabad.bits-pilani.ac.in}
\affiliation{Department of Mathematics, Birla Institute of Technology and
Science-Pilani,\\ Hyderabad Campus, Hyderabad-500078, India.}


\date{\today}
\begin{abstract}
In this letter, we explore the $f(Q,T_{\mu\nu}T^{\mu \nu})$ gravity theory, building upon the foundations laid by the $f(Q)$ and $f(Q,T)$ gravity theories. Here, $Q$ represents non-metricity and $T_{\mu\nu}$ stands for the energy-momentum tensor. The proposed action encompasses an arbitrary function of both non-metricity $Q$ and the square of the energy-momentum tensor, specifically $T^2=T_{\mu\nu}T^{\mu \nu}$. We find the analytical solution for the barotropic fluid case $p=\omega \rho$ for the model $f(Q, T_{\mu \nu}T^{\mu \nu})  = Q + \eta(T_{\mu \nu}T^{\mu \nu}) $. We constrain parameters of the solution $H(z)$ utilizing CC, BAO, and latest Pantheon+SH0ES samples with the help of Monte Carlo Markov Chain sampling technique along with Bayesian statistical analysis. Further, from the Om diagnostic test, we find that the assumed cosmological model favors the quintessence regime.


\end{abstract}

\maketitle

\section{Introduction}\label{sec1}
The advent of General Relativity (GR) in 1916, courtesy of Albert Einstein, marked a paradigm shift in our comprehension of gravity \cite{EIN}. Over the ensuing century, GR has withstood rigorous empirical scrutiny, vindicated by its accurate predictions: from Mercury's perihelion precession to the bending of starlight by the Sun, gravitational redshift, and even the monumental detection of gravitational waves \cite{BP} from cosmic cataclysms. Yet, amidst these triumphs, challenges have emerged, chiefly highlighted by the cosmic acceleration discovered in the early 21st century \cite{AG}. This revelation hinted at the limitations of GR on cosmological scales and underscored the pressing need for a more encompassing theory. Furthermore, GR's inability to reconcile with quantum mechanics leaves a conspicuous gap in our understanding, particularly when probing the fabric of spacetime at minuscule scales. Moreover, standard cosmology, rooted in GR, grapples with persistent puzzles: the enigmatic singularity problem, the confounding cosmological constant dilemma, and the perplexing cosmic coincidence quandary. These unresolved conundrums serve as poignant reminders that our current cosmological framework is far from complete. Thus, the past century has witnessed a surge of theoretical inquiry, driven by the imperative to reconcile observational evidence with theoretical predictions. Explorations into alternative gravity theories, quantum gravity frameworks, and novel cosmological paradigms represent the vanguard of this endeavor, aiming to furnish a more comprehensive elucidation of the cosmos.This motivation sparked huge theoretical investigations in understanding of our Universe  over the past century \cite{SN,SD,SC}. As quantum mechanics matured throughout this period, scientists, including Einstein himself, embarked on the quest to formulate a coherent theory of quantum gravity. This pursuit gave rise to various contenders such as string theory, loop quantum gravity theory, among others. However, despite their promise, none have yet achieved the status of being truly comprehensive.
To address the challenge of reconciling General Relativity's predictions with observations on cosmological scales, mainstream research has diverged into two distinct paths.\\
The initial avenue explores the concept of dark energy (DE), which theorizes that the universe's material composition can be depicted by an unusual fluid exerting negative pressure, consequently driving the observed acceleration of cosmic expansion. Numerous comprehensive reviews on dark energy can be found in literature. On the alternative path, scientists explore modified gravity theories. Here, the gravitational framework of General Relativity undergoes alterations to accommodate and explain the phenomenon of cosmic acceleration.Buchdahl introduced the $f(R)$ model as the pioneering modified gravity theory in 1970. In this theory, the conventional Einstein-Hilbert action's curvature scalar $R$ is substituted with an arbitrary function $f(R)$. Significant advancements in $f(R)$ gravity have been documented in \cite{PR,PR1,LA,YS}. A coupling between the matter sector and the gravity sector is contemplated through the function $f(R, T)$. Katirci and Kavuk proposed $f(R,T^2)$ in \cite{NK}, where $T^{2} = T_{\mu\nu}T^{\mu \nu}$ and $T_{\mu\nu}$ represents the matter energy-momentum tensor. Roshan and Shojai further delved into the theory, exploring the properties of the form $R + T^2$, termed as energy-momentum squared gravity (EMSG) \cite{MR}. Evolutionary alterations in gravitational theories often entail elaborating on the Einstein-Hilbert action within General Relativity, which relies on the curvature-based depiction of gravity. Nonetheless, an alluring alternative emerges when we broaden our scope to include the action stemming from the equivalent formulation of General Relativity that integrates torsion. Notably, Einstein also devised the "Teleparallel Equivalent of General Relativity" (TEGR).In this framework, the gravitational field is characterized by the torsion tensor rather than the curvature tensor.\\
In this article, our aim is to delve into a further extension of the symmetric teleparallel theory. Specifically, we are motivated to progress from $f(Q,T)$ to $f(Q,T^2)$ gravity, where $T^{2}$ is defined as $T_{\mu\nu}T^{\mu \nu}$, inspired by the extension of $f(R,T)$ gravity to $f(R,T^2)$. We shall refer to this extension as Energy-Momentum Squared Symmetric Teleparallel Gravity (EMSSTG), denoted by $f(Q, T^2)$. The gravitational action will be governed by an arbitrary function $f(Q, T^2)$ of $Q$ and $T^2$. Subsequently, by varying the action with respect to the metric, we can derive the field equations within a metric-affine formalism. These equations will serve as the foundation for exploring the cosmological evolution of the theory in depth. Investigating a specific toy model may yield valuable insights into effectively comprehending the dynamics of the theory.

\section{The Mathematical Formulation of  $f(Q,T_{\mu\nu}T^{\mu \nu})$  Theory}\label{sec2}
\justify
The generic action of the $f(Q,T_{\mu\nu}T^{\mu \nu})$ theory is given by \cite{PRR},
\begin{equation}\label{action}
S = \int \left[ \frac{1}{16\pi} f(Q,T^2) + L_m \right] \sqrt{-g} \, d^4x 
\end{equation}
Here, $g = \det(g_{\mu\nu})$   and 
 $T^{2} = T_{\mu\nu} T^{\mu \nu}$
 where $T_{\mu\nu}$ is the matter energy-momentum tensor.
The nonmetricity tensor $ Q_{\lambda \mu\nu}$ emerges as the result of the covariant derivative of the metric tensor concerning the Weyl-Cartan connection $ \bar{\Gamma}^{\lambda}_{\mu \nu}$. It can be represented as:

\begin{equation}
-\nabla_\lambda g_{\mu \nu} = Q_{\lambda \mu\nu} = -\frac{\partial g_{\mu \nu}}{\partial x^{\lambda}} + g_{\nu \sigma}\bar{\Gamma}^{\sigma}{\mu \lambda} + g_{\sigma \mu}\bar{\Gamma}^{\sigma}_{\nu \lambda}.
\end{equation}

The Christoffel symbol $\bar{\Gamma}^{\lambda}_{\mu \nu}$, the contortion tensor $C^{\lambda}_{\mu \nu}$, and the disformation tensor $L^{\lambda}_{\mu \nu}$ can be combined to form the Weyl-Cartan connection $ \bar{\Gamma}^{\lambda}_{\mu \nu}$ as:

\begin{equation}
\bar{\Gamma}^{\lambda}_{\mu \nu} = {\Gamma}^{\lambda}_{\mu \nu} + C^{\lambda}_{\mu \nu} + L^{\lambda}_{\mu \nu}
\end{equation}
In the provided equation, the initial term, referred to as the Christoffel symbol, essentially represents the Levi Civita connection associated with the metric tensor, denoted by $ g_{\mu _\nu} $. This connection serves as a fundamental concept in differential geometry, offering insights into the curvature and geometry of a given manifold  and is given by 
\begin{equation}
 {\Gamma}^{\lambda}_{\mu \nu} = \frac{1}{2}g^{\lambda \sigma}\bigg(\frac{\partial g_{\sigma \nu}}{\partial x^{\mu}} + \frac{\partial g_{\sigma \mu}}{\partial x^{\nu}} - \frac{\partial g_{\mu \nu}}{\partial x^{\sigma}} \bigg) 
\end{equation}
The contortion tensor $C^{\lambda}_{\mu \nu}$ is derived from torsion tensor $\bar{\Gamma}^{\lambda}_ {[\mu \nu]}$ which is given as : 
\begin{equation}
\bar{\Gamma}^{\lambda}_ {[\mu \nu]} = \frac{1}{2}\bigg(\bar{\Gamma}^{\lambda}_{\mu \nu} - \bar{\Gamma}^{\lambda}_{\nu \mu}\bigg)  
\end{equation}
By employing the provided definition, the contortion tensor is articulated as:
\begin{equation}
C^{\lambda}_{\mu \nu} = \bar{\Gamma}^{\lambda}_ {[\mu \nu]} + g^{\lambda \sigma}g_{\mu k}\bar{\Gamma}^{k}_ {[\nu \sigma]} + g^{\lambda \sigma}g_{\nu k}\bar{\Gamma}^{k}_ {[\mu \sigma]}
\end{equation}
The disformation tensor is articulated as:
\begin{equation*}
L^{\alpha}_{\beta \gamma} = -\frac{1}{2}g^{\alpha \lambda}(Q_{\gamma \beta \lambda} + Q_{\beta \lambda \gamma} -Q_{\lambda \gamma \beta})
\end{equation*}    
With this definition, the expression for the non-metricity tensor is obtained as: 
\begin{equation}
Q = -g^{\mu \nu}(L^{\alpha}_{\beta \mu} L^{\beta}_{\nu \alpha} - L^{\alpha}_{\beta \alpha} 	L^{\beta}_{\mu \nu})
\end{equation}
In the coincident gauge, where the covariant derivative simplifies to the partial derivative, the non-metricity invariant indeed aligns with the negative of the Einstein Hilbert Lagrangian.This choice of gauge is termed the coincident gauge and aligns seamlessly with symmetric teleparallel gravity.
Moreover the Weyl-Cartan torsion tensor $\tau^{\lambda}_{\mu \nu}$ is defined as :
\begin{equation}
\tau^{\lambda}_{\mu \nu} = \frac{1}{2}(\bar{\Gamma}^{\lambda}_{\mu \nu} - \bar{\Gamma}^{\lambda}_{\nu \mu})
\end{equation}
The Weyl-Cartan curvature tensor  is a fundamental geometric object in theories of gravity that incorporate both curvature and torsion is defined as : 
\begin{equation}
\bar{R}^{\lambda}_{\mu \nu \sigma} = \bar{\Gamma}^{\lambda}_{\mu \sigma,\nu} - \bar{\Gamma}^{\lambda}_{\mu \nu, \sigma} + \bar{\Gamma}^{\alpha}_{\mu \sigma}\bar{\Gamma}^{\lambda}_{\alpha \nu} -\bar{\Gamma}^{\alpha}_{\mu \nu}\bar{\Gamma}^{\lambda}_{\alpha \sigma}
\end{equation}
The trace of the energy-momentum tensor $T_{\mu\nu}$ is expressed as:
\begin{eqnarray}
T = g^{\mu\nu}T_{\mu\nu} = g_{\mu\nu}T^{\mu \nu} 
\end{eqnarray}
and the trace of  non-metricity tensor is given by,
\begin{eqnarray}
Q_{\beta} = Q^{\lambda}_{\beta \lambda} = \bar{Q}_{\beta} = Q^{\lambda}_{\beta \lambda} 	
\end{eqnarray}
The superpotential of the model is defined as 
\begin{eqnarray}
\hspace{-1.5cm} P^{\alpha}_{\mu \nu} &=& \frac{1}{4}\bigg[- Q^{\alpha}_{\mu \nu} + 2Q^{\alpha}_{(\mu \ \nu)} + Q^{\alpha}g_{\mu \nu} - \bar{Q}^{\alpha}g_{\mu \nu} - \delta^{\alpha}_{(_\mu Q _\nu}) \bigg]\\ &=& -\frac{1}{2}L_{\mu \nu}^{\alpha} + \frac{1}{4}\bigg( Q^{\alpha} - \bar{Q}^{\alpha} \bigg)g_{\mu\nu}  - \frac{1}{4}\delta^{\alpha}_{(_\mu Q _\nu})
\end{eqnarray}
Utilizing the superpotential in conjunction with the non-metricity tensor facilitates the derivation of the following relationship:
\begin{equation}
Q = -Q_{\lambda \alpha \beta} P^{\lambda \alpha \beta}     
\end{equation}
 We finally obtain the field equations of $f(Q,T^2)$ as follows:
\begin{small}
\begin{multline*}
-\frac{2}{\sqrt{-g}}\nabla_{\alpha}(f_Q\sqrt{-g}P^{\alpha}_{\mu \nu}) - \frac{1}{2}f(Q, T^2)g_{\mu \nu} + f_{T^2}\theta_{\mu \nu} \\
- f_{Q}(P_{\mu \alpha \beta}Q^{\alpha \beta}_{\nu} - 2Q^{\alpha \beta}_{\mu}P_{\alpha \beta \nu}) = 8\pi T_{\mu \nu}
\end{multline*}
\end{small}
Here, ${T}_{\mu\nu}$ is the stress-energy tensor defined as,
\begin{equation}\label{2h}
{T}_{\mu\nu} = \frac{-2}{\sqrt{-g}} \frac{\delta(\sqrt{-g}L_m)}{\delta g^{\mu\nu}}
\end{equation}
and 
\begin{equation}
\theta_{\mu \nu} = \frac{(\delta T_{\alpha \beta} T^{\alpha \beta})}{\delta g^{\mu\nu}}  
\end{equation}
Furthermore, the connection field equation that results from varying equation \eqref{action} is as follows:
\begin{equation}
\nabla_{\mu}\nabla_{\nu}(\sqrt{-g}f_QP^{\mu \nu}_{\alpha} + 4\pi H^{\mu \nu}_{\alpha}) = 0
\end{equation}
where,
\begin{equation}
H^{\alpha \beta}_{\rho} = \frac{\sqrt{-g}}{16 \pi}f_{T^2}\frac{\delta T^2}{\delta T^{\rho}_{\alpha \beta}} + \frac{\delta \sqrt{-g}L_{m}}{\delta T^{\rho}_{\alpha \beta}}
\end{equation}

\section{COSMOLOGY IN EMSSTG  } \label{sec3}

We commence with the following homogeneous and isotropic flat Friedmann-Lemaitre-Robertson-Walker (FLRW) line element, presented in Cartesian coordinates:

\begin{equation}\label{3a}
ds^2= -dt^2 + a^2(t)[dx^2+dy^2+dz^2]
\end{equation}
where the cosmological scale factor, $a(t)$, represents the universe's expansion throughout time.
The non-metricity scalar $Q$ for the metric \eqref{3a} is given as
\begin{equation}\label{3b}
 Q= 6H^2  
\end{equation}

The energy-momentum tensor $T_{\mu\nu}$ for the matter, which is treated as a perfect fluid, is given as follows:

\begin{equation}
\label{9}
T_{\mu\nu} = (p + \rho) u_{\mu} u_{\nu} + p g_{\mu\nu},
\end{equation}

where $p$ and $\rho$ represent the pressure and energy density of the perfect fluid, respectively, and $u_{\mu}$ denotes a four-velocity vector. The Friedmann like equations for the line element \eqref{3a} is given by,

\begin{equation}
6f_QH^2 - \frac{1}{2}f(Q, T^2) = 8\pi \rho + f_{T^2}(\rho + 4p\rho +3p^2)
\end{equation}
\begin{equation}
6f_QH^2 - \frac{1}{2}f(Q, T^2) -2(\dot{f}_QH + f_Q\dot{H}) = - 8\pi p 
\end{equation}\\
where $(.)$ represents the derivate with respect to time.

We consider the following $f(Q, T^2)$ model based on the specific coupling nature between $Q$ and $T^2$ as follows,
\begin{equation}
f(Q, T_{\mu \nu}T^{\mu \nu}) = f(Q, T^2) = Q + \eta(T_{\mu \nu}T^{\mu \nu}) = Q + \eta(T^2)
\end{equation}

We assume the following barotropic equation of state for a fluid that typically relates pressure $(p)$ to density $(\rho)$ in a way that depends only on the local density,  
\begin{equation}
p=\omega\rho.
\end{equation}

The corresponding Friedman equation becomes,
\begin{equation}
3H^2 = 8\pi \rho +\eta [\frac{3}{2}(1+3\omega^{2})+4\omega]\rho^{2}
\end{equation}
\begin{equation}
3H^2 - 2\dot{H} = \frac{\eta}{2}({\rho}^2 + 3p^2) - 8\pi p
\end{equation}
where $\omega$ is the equation of state parameter.
    
The continuity equation for the assumed $f(Q, T^2)$ model is given as \cite{PRR}, 

\begin{multline}  
\dot{\rho}+3 H(\rho+p) \\ =\frac{3 H\left[18 H^{2}-4 \dot{H}+\rho(48 \omega \pi+\eta(1+\omega(14+\omega(19+18 \omega))) \rho)\right]}{16 \pi+2 \eta(3+\omega(8+9 \omega)) \rho}
\end{multline}

Using above equations we solve for $H(z)$ and get,
\begin{small}
\begin{equation}
H^2(z)=H_0^2(1+z)^{3(1+\omega)}[1+ \eta \Omega_0 ^{2} \{ \frac{3}{2}(1+3\omega^2)+4\omega \}
\{(1+z)^{3(1+\omega)}-1 \}]    
\end{equation}    
\end{small}

\section{Observational constraints and Methodology}
In this section, we'll employ a statistical analysis using the Monte Carlo Markov Chain (MCMC) method. Our objective is to evaluate the effectiveness of a model by comparing its predictions with cosmic observations. Specifically, we'll assess the model's viability by examining its consistency with Baryon Acoustic Oscillation (BAO) data and Observational Hubble data (H(z)).The Monte Carlo Markov Chain (MCMC) method holds a pivotal role in cosmological research, widely utilized for traversing parameter spaces and deriving associated probability distributions \cite{Almada/2019}. Fundamentally, MCMC involves the creation of a Markov chain, methodically exploring the parameter space by sampling according to a predefined probability distribution. The chain progresses as a series of parameter values, with each derived from the preceding value according to predetermined transition rules governed by a proposal distribution. This distribution introduces new parameter values, and their acceptance depends on their posterior probability, which incorporates both observational data and prior probability functions. Once the chain reaches convergence, an estimation of the posterior distribution for the parameters becomes achievable by tallying the frequency of parameter values within the chain. Consequently, this posterior distribution aids in determining the optimal parameter values and their associated uncertainties, thereby facilitating predictions for various observables.

\subsection{Cosmic chronometer data}
We employ Hubble parameter measurements derived through the application of the differential age method, commonly referred to as cosmic chronometer (CC) data. Our analysis encompasses 31 meticulously compiled data points. \cite{Moresco/2015}.
The $\chi^{2}$ function is given as 
\begin{equation}
\chi^2_{CC}= \sum_{i=1}^{31} \frac{\left[H(z_{i})-H_{obs}(z_{i})\right]^2}{\sigma(z_{i})^{2}},
\end{equation}
where the observational error is $\sigma(z_{i})$ and the observed value of $H(z)$ is $H_{obs}$.

\subsection{Pantheon+SH0ES data}
The latest SN compilation provides a comprehensive collection of 1701 samples, presenting a wealth of cosmological insights. Covering a broad spectrum of redshifts ranging from 0.001 to 2.3, these data points offer researchers a robust foundation for investigating the dynamics of cosmic expansion over an extensive temporal range. Going beyond previous compilations of Type Ia supernovae (SNIa), the Pantheon+ dataset incorporates the latest observations, enhancing its utility and relevance in cosmological studies. SNIa, renowned for their consistent brightness, serve as dependable standard candles, facilitating the estimation of relative distances across the universe via the distance modulus technique. Over recent years, several compilations of Type Ia supernova data, such as Union, \cite{R30}, Union2 \cite{R31}, Union2.1 \cite{R32}, JLA \cite{R33}, Pantheon \cite{R34}, and the latest addition, Pantheon+ \cite{R35}. The corresponding $\chi^2$ function reads as,
\begin{equation}\label{4c}
\chi^2_{SN}=  D^T C^{-1}_{SN} D,
\end{equation}
In this context, $C_{SN}$ represents the covariance matrix associated with the Pantheon+ samples, encompassing both statistical and systematic uncertainties. Additionally, the vector $D$ is defined as $D = m_{Bi} - M - \mu^{th}(z_i)$, where $m_{Bi}$ represents the apparent magnitude and $M$ denotes the absolute magnitude. Furthermore, the expression $\mu^{th}(z_i)$ represents the theoretical distance modulus of the Modified Gravity (MOG) model, which can be estimated using the following relation:
\begin{equation}\label{4d}
\mu^{th}(z_i)= 5log_{10} \left[ \frac{D_{L}(z_i)}{1 Mpc}  \right]+25, 
\end{equation}
In this context, $D_{L}(z)$ denotes the luminosity distance associated with the provided Modified Gravity (MOG) model and can be computed using the following expression:

\begin{equation}\label{4e}
D_{L}(z)= c(1+z) \int_{0}^{z} \frac{ dx}{H(x,\theta)}
\end{equation}

Here, $\theta$ represents the typical parameter space, and $c$ stands for the speed of light. The integral spans from 0 to $z$, where $z$ represents the redshift, and $H(x,\theta)$ represents the Hubble parameter at a given redshift $x$ and within the parameter space $\theta$.
Unlike the Pantheon dataset, the Pantheon+ compilation effectively addresses the degeneracy between the parameters $H_0$ and $M$ by redefining the vector $D$ as follows:
\begin{equation}\label{4f}
\bar{D} = \begin{cases}
     m_{Bi}-M-\mu_i^{Ceph} & i \in \text{Cepheid hosts} \\
     m_{Bi}-M-\mu^{th}(z_i) & \text{otherwise}
    \end{cases}   
\end{equation}
where $\mu_i^{Ceph}$ independently estimated using Cepheid calibrators. Hence, the relation \eqref{4c} becomes $\chi^2_{SN}=  \bar{D}^T C^{-1}_{SN} \bar{D} $.
\subsection{BAO}
Baryonic Acoustic Oscillations (BAO) serve as a vital tool in cosmology, enabling us to probe the vast structure of the Universe on a grand scale. These fluctuations originate from acoustic waves that propagated through the early Universe, causing the compression of baryonic matter and radiation within the photon-baryon fluid. This compression creates a unique peak in the correlation function of galaxies or quasars, providing a consistent ruler for measuring cosmic distances. The characteristic size of the BAO peak is determined by the sound horizon at the time of recombination, which depends on factors such as the density of baryons and the temperature of the cosmic microwave background.On large angular scales, baryon acoustic oscillations (BAO) occur as separate peaks and are thought to be pressure waves caused by cosmic perturbations in the baryon-photon plasma during the recombination era (BOSS) \cite{Blake/2011,Percival/2010}. The expressions utilized for non-correlated BAO data are as follows,

\begin{equation}
\chi^2_{BAO/non cov }= \sum_{i=1}^{26} \frac{\left[H_{th}(z_{i,\vartheta})-H_{obs}^{BAO}(z_{i})\right]^2}{\sigma^{2}(z_{i})},
\end{equation}
In these expressions, $H_{\text{th}}$ denotes the theoretical values of the Hubble parameter for a particular model characterized by model parameters $\vartheta$. Conversely, $H_{\text{obs}}^{\text{BAO}}$ corresponds to the observed Hubble parameter acquired through the BAO method, while ${\sigma_H}$ represents the error associated with the observed values of $H^{\text{BAO}}$. For the correlated BAO samples, the following expressions are utilized,

\begin{eqnarray}
d_{A}(z) &=& c \int_{0}^{z} \frac{dz'}{H(z')},\\
D_{v}(z) &=& \left[\frac{d_{A}(z)^2 c z }{H(z)}\right]^{1/3},\\
\chi_{BAO/cov}^2 &=& X^{T} C^{-1} X. 
\end{eqnarray}
The comoving angular diameter distance is denoted by $d_{A}(z)$, the dilation scale by $D_{v}(z)$, and the covariance matrix by $C$. \cite{Giostri/2012}.
Hence the total chi-square function for BAO samples is defined as :

\begin{equation}\label{4f}
\chi^2_{BAO}= \chi^2_{BAO/non cov}+\chi^2_{BAO/cov} 
\end{equation}
\begin{widetext}

The contour plot for the assumed model corresponding to free parameters within the $1\sigma-3\sigma$ confidence interval using CC, CC+BAO, CC+SN, and CC+BAO+SN samples presented in the Fig \eqref{f1}. The obtained parameters constraint are listed in the Table \eqref{table1}.

\centering
\begin{figure}[H]
\includegraphics[scale=0.8]{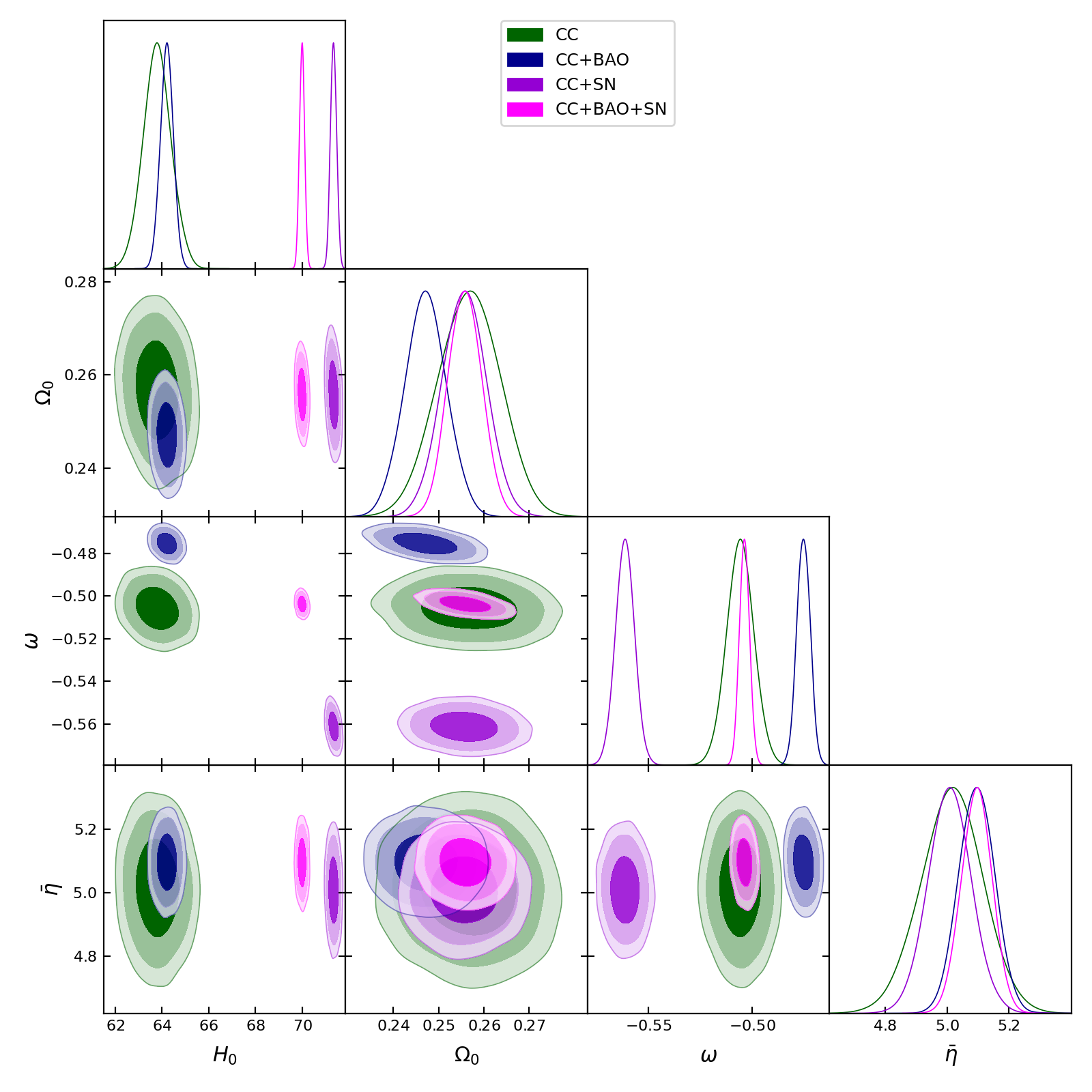}
\caption{The contour plot for the assumed model corresponding to free parameters within the $1\sigma-3\sigma$ confidence interval using CC, CC+BAO, CC+SN, and CC+BAO+SN samples.}\label{f1}
\end{figure}

\end{widetext}

\begin{table*}
\begin{center}
 \caption{Best-fit values of model parameters determined  from observational datasets}
 
    \label{table1}
\begin{tabular}{|l|c|c|c|c|}
\hline 
Datasets              & $CC$ & $CC+BAO$ & $CC+SN$ & $CC+SN+BAO$ \\\hline

$H_0$           & $64^{+0.59}_{-0.59}$  & $64^{0.27}_{-0.28}$ &  $71^{+0.13}_{-0.13}$ & $70^{+0.11}_{-0.11}$ \\
\hline

$\Omega_0$           & $0.26^{+0.0009}_{-0.0071}$  & $0.25^{0.0046}_{-0.0045}$ &  $0.26^{+0.0048}_{-0.0049}$ & $0.26^{+0.00038}_{-0.0037}$\\
\hline

$\omega$           & $-0.51^{+0.0065}_{-0.0065}$  & $-0.48^{0.0032}_{-0.0032}$ &  $-0.56^{+0.0045}_{-0.0045}$ & $-0.5^{+0.0024}_{-0.0024}$\\
\hline

$\eta$           & $5^{+0.098}_{-0.1}$  & $5.1^{0.058}_{-0.058}$ &  $5^{+0.068}_{0.069}$ & $5.1^{+0.049}_{-0.05}$\\
\hline

$q_0$ & $-0.108$  & $-0.066$ &  $-0.19$ & $-0.088$\\ 

\hline

$z_t$ & $0.646$  & $0.354$ &  $1.65$ & $0.491$\\ 
\hline

$\omega_0$ & $-0.405$  & $-0.377$ &  $-0.46$ & $-0.392$\\
\hline

$\chi^2 _{min}(model) $           & $19.448$  & $42.933$ &  $1621.46$ & $1644.631$\\
\hline

$\chi^2 _{min}(\Lambda CDM) $           & $26.597$  & $55.926$ &  $1640.198$ & $1669.527$\\

\hline

$AIC(model) $           & $30.597$  & $59.866$ &  $1644.198$ & $1674.257$\\
\hline

$AIC (\Lambda CDM) $           & $27.448$  & $50.933$ &  $1629.146$ & $1653.631$\\

\hline

$\Delta AIC $           & $3.149$  & $8.933$ &  $15.052$ & $20.626$\\
\hline

\end{tabular}
\end{center}
\end{table*}

\section{Evolutionary Behavior of Cosmological Parameters}
\begin{figure}[H]
\centering
\includegraphics[scale=0.4]{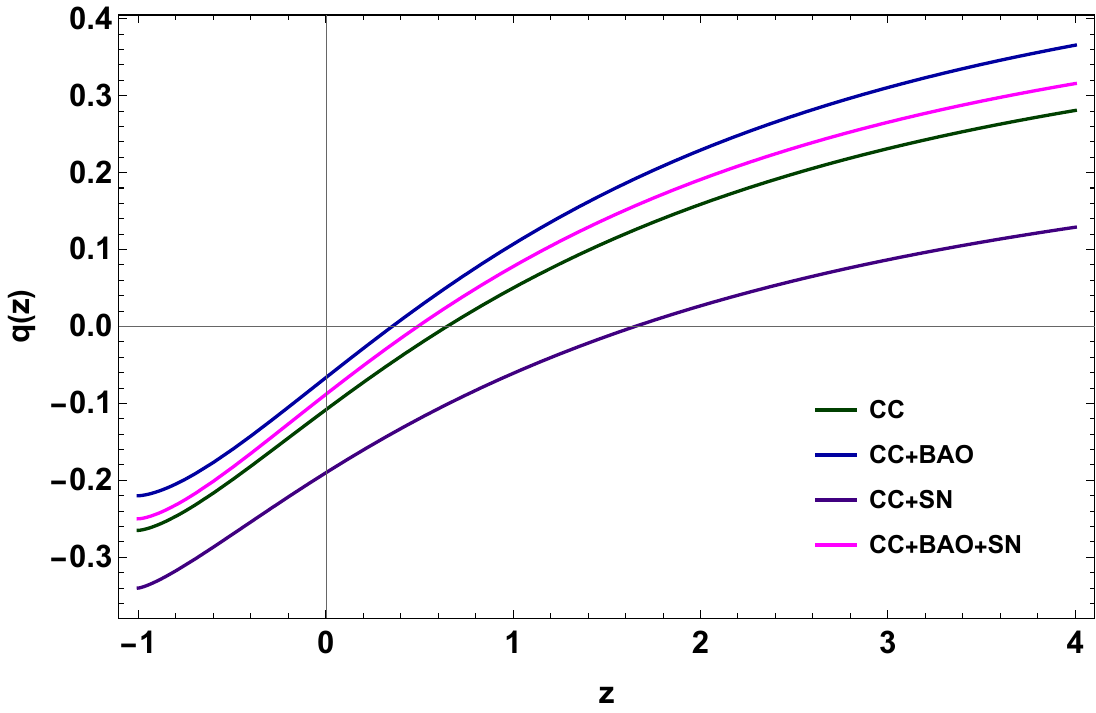}
\caption{Variation of the deceleration parameter $q$ as a function of the redshift $z$ for different data sets.}
\label{f2}
\end{figure}

\begin{figure}[H]
\centering
\includegraphics[scale=0.52]{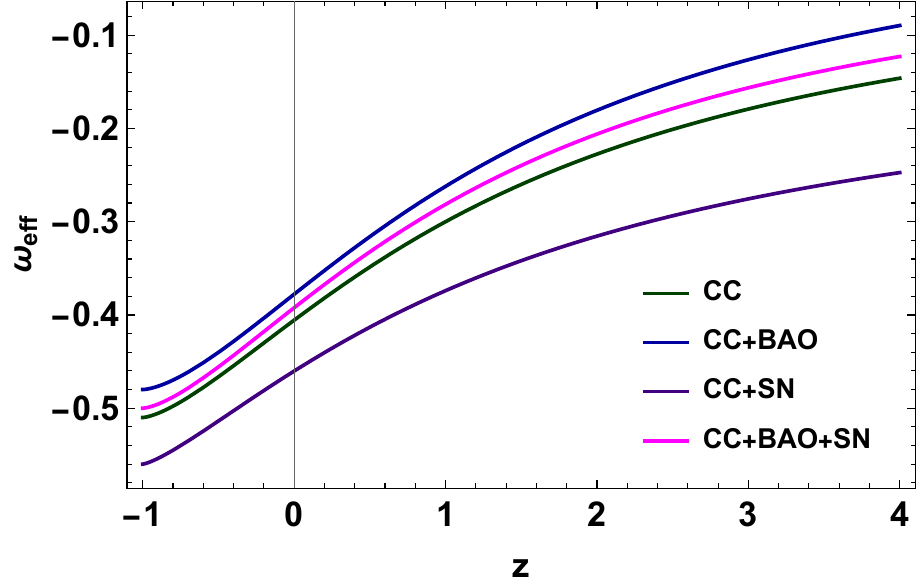}
\caption{Variation of the effective EoS parameter $\omega_{eff}$ as a function of the redshift $z$ for different data sets.}
\label{f2a}
\end{figure}

\begin{figure}[H]
\centering
\includegraphics[scale=0.6]{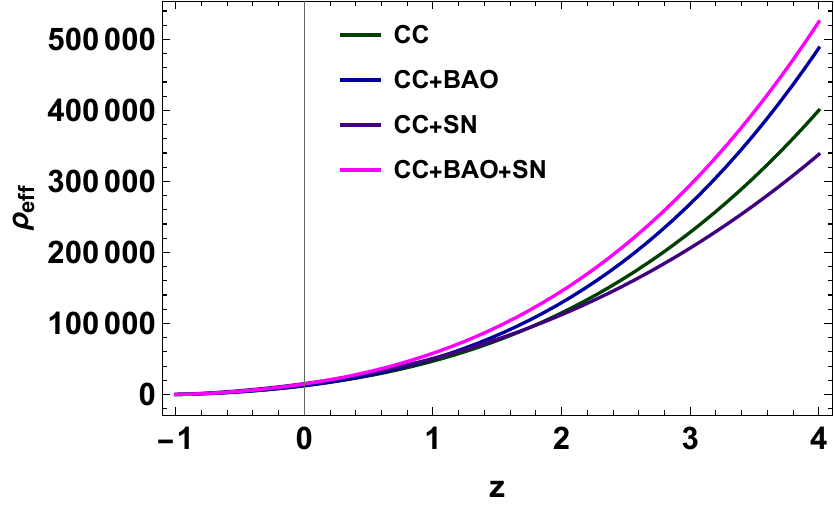}
\caption{Variation of the effective energy density $\rho_{eff}$ as a function of the redshift $z$ for different data sets.}
\label{f2b}
\end{figure}

\justifying
The visualizations shown below vividly illustrate how the dynamics of the universe can exhibit remarkable intricacies, contingent upon the specific values of the parameters involved. Figure \ref{f2} illustrates the universe's trajectory, commencing with a decelerating phase ($q>0$) before transitioning to an accelerating phase ($q<0$) following a transition redshift denoted as $z_{t}$. The deceleration parameter, denoted as $q$, is calculated using the expression $q= -\frac{\dot{H}}{H^2}-1$. This evolutionary pattern aligns with contemporary understanding of the universe's behavior, characterized by three distinct stages: an initial decelerating phase, a subsequent period of accelerating expansion, and a late-time acceleration phase. Remarkably, our results demonstrate that the present-day value of the deceleration parameter ($q_{0}$) depicts the acceleration phase \cite{Almada/2019,Basilakos/2012} and the transition redshift ($z_{t}$) \cite{Garza/2019,Jesus/2020} align well with observations from taken dataset, listed in the Table \eqref{table1}. Further, the same result is reflected in the behavior of the effective equation of state parameter defined by $\omega_{eff}= -\frac{\dot{2H}}{3H^2}-1$, presented in the Figure \ref{f2a}. Moreover, the effective matter-energy density show expected positive behavior in the entire domain of redshift, presented in the Figure \ref{f2b}. Note that, we observe that the trajectories of the cosmological evolutionary parameters corresponding to CC+SN samples is much deviated in comparison to other datset combinations. The underlying root cause of this deviation is the nature of datasets, such as BAO and CC datasets are more sensitive to the early universe and can describe the proper transition epoch, whereas the SN datasets is concentrated at lower redshifts, mostly less than one, therefore it more focused on the present cosmic acceleration rather than the full history of expansion. Therefore the presence of SN data points in CC+SN samples favors a high transition value, whereas the CC and BAO combination provides a true transition value. For instance, one must check the reference \cite{CPS,R29,R30} to see how the SN datapoints prefers a high transition redshift, generally greater than one, whereas the observations of BAO and CC prefer a low transition redshift, generally less than one, and thus this will lead to a discrepancy in the $H_0$, $\Omega_0$, as well as $z_t$ value due to the nature of underlying datasests and its measurement techniques.\\
A simple test technique that uses only the cosmic scale factor's first-order derivative is the Om diagnostic. Its equation in the case of a spatially flat universe is as follows \cite{R38}:

\begin{equation}\label{5b}
Om(z)= \frac{\big(\frac{H(z)}{H_0}\big)^2-1}{(1+z)^3-1}
\end{equation}

\begin{figure}[H]
\centering
\includegraphics[scale=0.4]{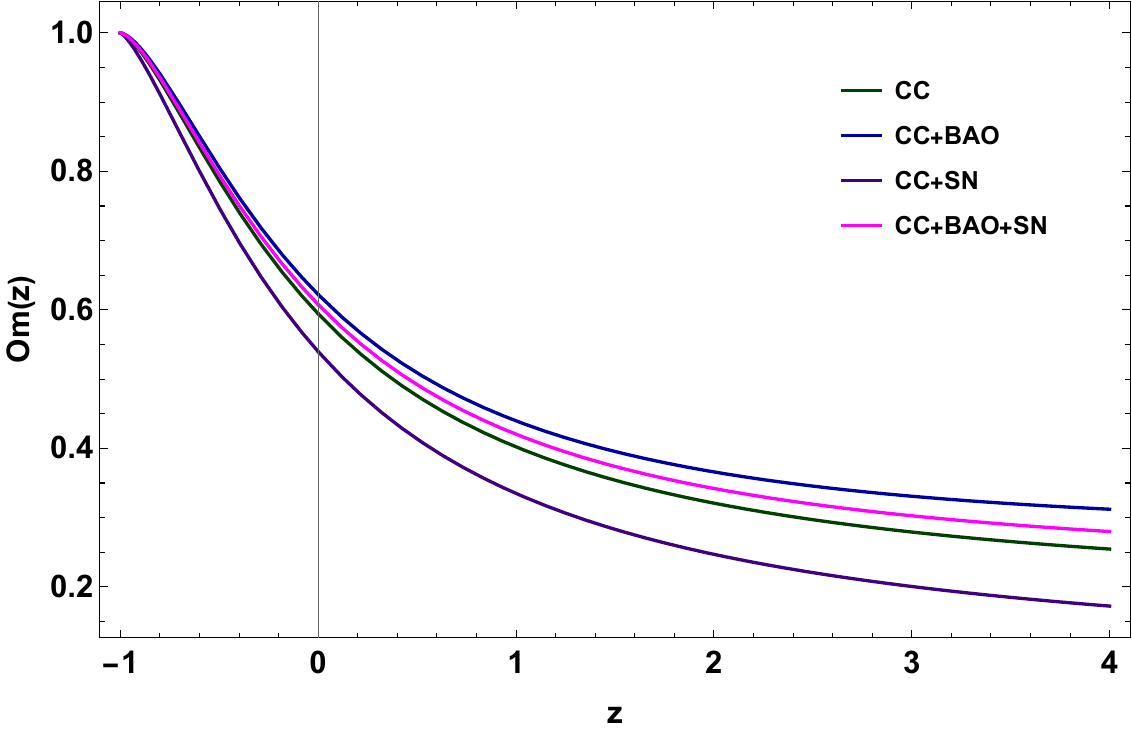}
\caption{Variation of Om diagnostic parameter as a function of the redshift $z$ for different data sets.}
\label{f3}
\end{figure}

The descending slope of the $Om(z)$ curve indicates quintessence-like behavior, while an ascending slope corresponds to phantom behavior. Conversely, a constant $Om(z)$ signifies the characteristics of the $\Lambda$CDM model. From the behavior of Om diagnostic parameter presented in the Fig \eqref{f3}, it can be inferred that our cosmological framework exhibits quintessence-like behavior.

\section{Conclusion}\label{sec4}
\newpage

\justify In this study, we introduced a further extension of symmetric teleparallel gravity by broadening the gravity Lagrangian with an arbitrary function of $f(Q,T_{\mu\nu}T^{\mu \nu})$. We derived the FLRW equations for a flat, homogeneous, and isotropic spacetime. To deepen our understanding of the cosmological framework within this theory, we found the analytical solution for the barotropic fluid case $p=\omega \rho$ for the model $f(Q, T_{\mu \nu}T^{\mu \nu})  = Q + \eta(T_{\mu \nu}T^{\mu \nu})$. Further, we constrained parameters of the obtained solution $H(z)$ utilizing CC, BAO, and latest SN samples with the help of Monte Carlo Markov Chain sampling technique along with Bayesian statistical analysis. The obtained constraints on the parameters of considered cosmological settings are listed in the Table \eqref{table1}, along with the corresponding contour plots depicting the parameter correlation, in the Fig \eqref{f1}. In addition, we analyzed the behavior of deceleration parameter in the Fig \eqref{f2} depicting the observed accelerating phenomenon with the transition epoch. The present value of the deceleration parameter along with the transition redshift is listed in the Table \eqref{table1}. Further, the same result is reflected in the behavior of the effective equation of state parameter defined by $\omega_{eff}= -\frac{\dot{2H}}{3H^2}-1$, presented in the Figure \ref{f2a}. Moreover, the effective matter-energy density show expected positive behavior in the entire domain of redshift, presented in the Figure \ref{f2b}.  Moreover, we employed the Om diagnostic test to assess the behavior of supporting dark energy. We found that the behavior of Om diagnostic parameter presented in the Fig \eqref{f3} favors the quintessence type dark energy model. Thus, our investigation successfully describe late time expansion phase of the universe. However, we would like to note that as the square gravity is inherently dependent on the choice of Lagrangian density ($\mathcal{L}$), it can not accommodate scalar fields like inflation fields as $P_{\phi}=\frac{1}{2}\dot{\phi}^2-V(\phi)$ that is very different from $\mathcal{L}_{fluid}=P$. This, incorporating the scalar field into $T$ and $T^2$ gravity, come along with more complex issues as discussed the same in case of $f(R,T)$ gravity \cite{THH}. \\

\textbf{Data availability:} There are no new data associated with this article.\\

\section*{Acknowledgments} \label{sec7}
 Aaqid Bhat expresses gratitude to the BITS-Pilani, Hyderabad campus, India, for granting him a Junior Research Fellowship. PKS acknowledges Science and Engineering Research Board, Department of Science and Technology, Government of India for financial support to carry out Research project No.: CRG/2022/001847 and IUCAA, Pune, India for providing support through the visiting Associateship program.


\end{document}